\documentclass[prl,twocolumn]{revtex4-1}

\usepackage{amsmath}    
\usepackage{graphicx}   
\usepackage{subfigure}
\usepackage{verbatim}   
\usepackage{color}      
\usepackage{subfigure}  

\begin{document}
\title{Order parameter prediction from molecular dynamics simulations in proteins}
\author{Juan R. Perilla}
\author{Thomas B. Woolf}
\affiliation{Department of Biophysics and Biophysical Chemistry, Johns Hopkins University, Baltimore, Maryland 21205}
\date{\today}

\begin{abstract}
A molecular understanding of how protein function is related to protein structure will require an ability to understand large conformational changes between multiple states.  Unfortunately these states are often separated by high free energy barriers and within a complex energy landscape. This makes it very difficult to reliably connect, for example by all-atom molecular dynamics calculations, the states, their energies and the pathways between them.  A major issue needed to improve sampling on the intermediate states is an order parameter -- a reduced descriptor for the major subset of degrees of freedom -- that can be used to aid sampling for the large conformational change.  We present a novel way to combine information from molecular dynamics using non-linear time series and dimensionality reduction, in order to quantitatively determine an order parameter connecting two large-scale conformationally distinct protein states.  This new method suggests an implementation for molecular dynamics calculations that may dramatically enhance sampling of intermediate states.
\end{abstract}
\maketitle

%


\section{Introduction}
Proteins represent complex dynamical systems with many different stable states. Sampling on these systems has been termed the multiple time scale problem \cite{Berne1985} with many different wells and barrier heights being found. Despite considerable effort, there is currently no rapid way to determine the range of stable states from a single structure or from sequence alone.  Because biological function is intrinsically linked to the large scale conformational change of proteins, an improved understanding of how conformational change in the complex energy landscape of the protein is determined will have an immediate impact on biological and physical questions. 


One method that has been used exhaustively for attempting to determine reduced descriptors for protein motions has been through the concepts of normal mode analysis \cite{tamartb, zhenggoenm, vanAalten:Proteins:1995,Hayward:Proteins:1997}.  If the potential function driving the energies was fully harmonic, then the solution would be analytic and all the stable states and their motions would be immediately determined.  But, with thermal motions driving non-harmonic interactions such as van der Waals and electrostatic forces, the motions are controlled by a very complex, but physically defined, energy surface.  The current paradigm for interpreting these motions is to run the longest possible molecular dynamics calculations and then to use the fluctuations seen in the trajectory to define a set of effective normal modes.  These normal modes have been called the principal normal modes when they are restricted to the lowest collective degrees of freedom.  Several groups have suggested that following these normal modes could be used as an effective order parameter for controlling large conformational change\cite{vanAalten:Proteins:1995,Skjaerven:PlosComputBiol:2011,Miloshevsky:BiophysJ:2010}.  Nevertheless, the approach has not led to great success, with the modes often leading to \emph{blind alleys} in the surface that don't reflect large conformational change\cite{Balsera_Wriggers_Oono_Schulten_1996, Ma2005373, Petrone:BiophysJ:2006, Ramanathan:JPhysChemB:2009}.

To improve on the sampling of large-scale conformational change there have been many methods proposed.  The most well known contemporary method has been called transition path sampling and uses a small number of conformations along a candidate transition pathway along with a Monte Carlo move set to try and anneal an optimal prediction of intermediates in a conformationally changing system \cite{dellago:1998, Bolhuis:ProcNatlAcadSciUSA:2000}.  Alternative methods have used the RMS differences between states as an order parameter to control change \cite{Maragakis:JMolBiol:2005}, directly adding a new force that biases motion along the RMS gradient.  We have developed a method, called dynamic importance sampling (DIMS)\cite{woolf:1998,zuckerman:physreve,Zuckerman1999,zuckerman:2002,Perilla2010} that uses concepts from stochastic differential equations \cite{wagner:1987} to create a family of independent transitions that together define the likelihood of different pathways and the kinetics of the transition with sufficient sampling.  However, similar to the RMS based methods, the DIMS method requires a progress variable to gauge progress of the transitions and to create the biasing and its correction for an unbiased estimate of pathways, kinetics and states. 

In this contribution we describe the use of effective transfer entropy for the determination of a reduced set of degrees of freedom that can be used to define order parameters behind large scale conformational change.  Our approach combines insights from the physics of non-linear time-series analysis, dimensionality reduction, and the chemical physics of protein motions on a complex energy surface to enable the dynamics of the complex system to define an order parameter candidate.  This improves on other methods for the determination of order parameters where the candidate order parameter was inferred from empirical analysis of the static structure or simply assumed to correlate with the RMS between two different states.  In the calculations to follow we mainly use the receiver domain of nitrogen regulatory protein C (NtrC) (Fig. \ref{fig:pca}), in addition, we have performed steered molecular dynamics and checks of the implementation on the glucose-galactose binding protein (GGBP). 





\section{Principal component analysis}

The method of principal component analysis has been used in the analysis of protein motions for many years. This approach depends on the determination of a set of effective harmonic modes that define the complex motions that have been seen in the dynamics.  While the initial excitement over the method as a way to obtain longer time-scales seems to have faded, there remains much effort to use this approach as a tool for the analysis of conformational change. Recent work has suggested that this approach may lead to significant systematic error when there are multiple stable states separated by a large barrier.  

To compute the normal modes a molecular dynamics (MD) trajectory is used along with the determination of the average fluctuations in the simulation.  Then, from the MD trajectory $\vec{q}(t)=(q_1(t),q_2(t),\ldots,q_{3N}(t)))$ of a protein with $N$ atoms, the correlation matrix $\mathbf{\sigma}$ is built as follows:

\begin{equation}
\mathbf{\sigma}=\langle 
(\vec{q}(t) - \langle \vec{q}(t) \rangle)
(\vec{q}(t) - \langle \vec{q}(t) \rangle)
 \rangle. 
\label{eq:cormat}
\end{equation}
 
Where the brackets ($\langle \ldots \rangle$) denote time averages. The orthonormal basis vectors (principal components/PC) $\vec{\eta}_\alpha$ are determined by the eigenvalue problem $\lambda_\alpha\vec{\eta}_\alpha=\mathbf{\sigma}$. 

\begin{figure}
\centering
\includegraphics[width=0.9\columnwidth]{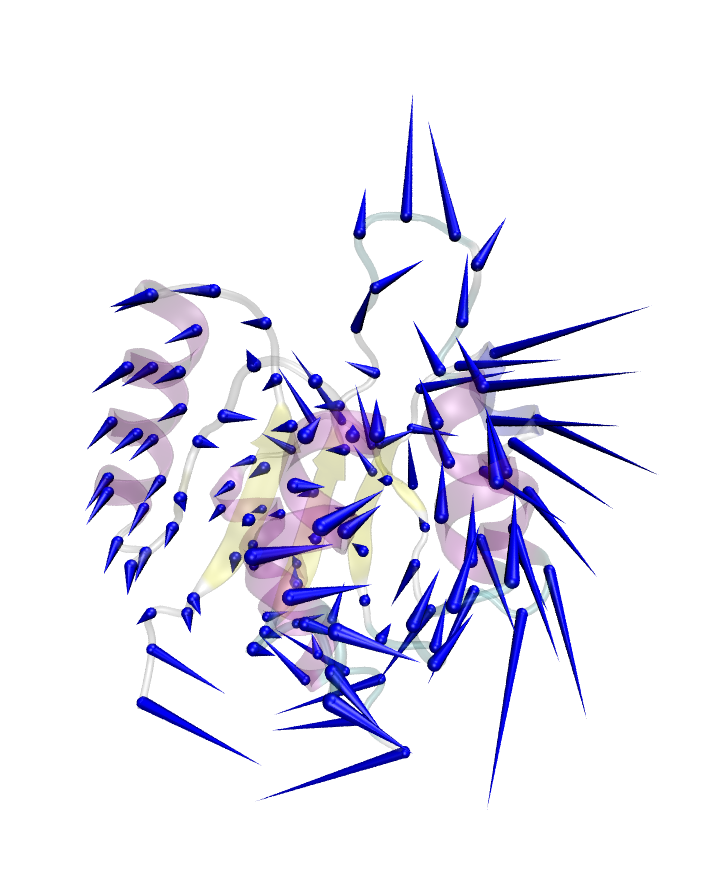}
\caption{Lowest frequency mode from PCA analysis for the inactive-state of NtrC .\label{fig:pca}}
\end{figure}

The lowest frequency modes from PCA are normally associated with slow, collective motions and have been used to try and predict intermediate states. Figure \ref{fig:pca} depicts the lowest frequency mode obtained by applying equation (\ref{eq:cormat}) and, solving the eigenvalue problem for our 600 ns trajectory of NtrC. On this plot the porcupine spines are located at the C${}_\alpha$ atoms and their magnitude and direction shows the type of motion involved in the mode.

For a given mode $\alpha$, the involvement coefficients (IC) are defined as:

\begin{equation}
\nu_\alpha=\| \vec{\eta}_\alpha \cdot (\hat{q}^A - \hat{q}^B ) \|,
\label{eq:involvement}
\end{equation}

where $\hat{q}^{A,B}$ indicates the set of normalized coordinates ($\hat{q}^{A,B}\cdot\hat{q}^{A,B}=1$) that represent the active-state and inactive-state conformations, respectively. Therefore, the ICs measure the amount of overlap between a PC and the direction defined by the displacement vector between structures. As mentioned in the case of hinge-bending motions, PCA shows higher values for the ICs compared to those from more complex motions. For instance, in the case of Adenalyte Kinase (AdK), the ICs for the first two modes are 0.49 and 0.63, respectively. In the case of NtrC the ICs are much lower (Fig. \ref{fig:involvement}), in consequence the direction of the first PCs of both stable states are not pointing directly towards the other end state. In contrast, in another study, by using an empirical order parameter (based on observations of both stable structures) yields higher ICs values \cite{Lei2009}. These order parameters involve only localized regions of the system and are proposed in an orderly series of events. This result would imply that the PCs are not pointing directly towards the other end state but towards a different intermediate state. 

One of the ideas behind an order parameter is that a few degrees of freedom dominate part of or the entire transition, while the rest of the system would follow. Therefore finding an order parameter is equivalent to locating such leading modes. In this paper we use an information theoretical approach to identify the leading modes by measuring the transfer entropy between pairs of residues. The more dominant residues are those that transfer the largest amount of entropy to the rest of the system. 


\section{Information flow in proteins}

The networks of interactions between atoms and residues define the web of dependencies and patterns of dynamic coupling between domains in a protein, characterized by the directed flow of information spanning multiple spatial and temporal scales. An initial application of transfer entropy to DNA binding proteins was the first to apply the asymmetry of information transfer to protein molecular motions\cite{Kamberaj20091747}.   Let $X$ be the time series for the center of mass of the $i$th residue and, $p(X)$ its probability distribution. Therefore it is possible to measure the average number of bits needed to optimally encode independent draws by using the Shannon entropy $H_{X}=-\sum_x p_x \log p(x)$\cite{Reza_1994,Shannon_1948}, where the sum extends over all the states that $X$ can reach. 

\subsection{Transfer entropies}

For a residue $j \neq i$ with a center of mass $Y$ and, probability distribution $p(Y)$; one could say that its trajectory is independent of that of residue $i$ if

\begin{equation}
p(y_{n+1}|y_n)=p(y_{n+1}|y_n,x_n),
\label{eq:indep}
\end{equation}

where $p(y_{n+1}|y_n)$ is the conditional probability to find residue $j$ at state $y_{n+1}$ given its past $y_{n},\ldots,y_{1}$ and, $p(y_{n+1}|y_n,x_n)$ is the conditional probability to find residue $j$ at state $y_{n+1}$ given the past of both $i$ and $j$. In the case where there is not a flux of information from $X$ to $Y$ then equation (\ref{eq:indep}) is correct. On the other hand, in the event that there is flux of information in any direction, the divergence from correctness of equation (\ref{eq:indep}) can be quantified by the Kullback-Leibler entropy \cite{Kullback_Leibler_1951} hence defining the transfer entropy\cite{Schreiber2000}: 

\begin{equation}
T_{X\to Y}=\sum p(y_{n+1},y_n,x_n) \log \frac{p(y_{n+1} | y_n, x_n)}{p(y_{n+1}|y_n)}.
\label{eq:trinf}
\end{equation}

The transfer entropy between $i$ and $j$ is minimum and equal to zero when the two residues are independent and there is a maximum and equal to the entropy rate:

\begin{equation}
h_Y = - \sum p(y_{n+1},y_n) \log p(y_{n+1}|y_n),
\label{eq:hrate}
\end{equation}

when the residues are completely coupled. In order to minimize artifacts within the time series we use the normalized effective transfer entropy given by \cite{Gourevitch2007,Marschinski2002}:

\begin{equation}
T^E_{X\to Y}= \frac{1}{h_Y} \left(T_{X\to Y} - \frac{1}{N_{\mathrm{trials}}}\sum_{n=1}^{N_{\mathrm{trials}}} T_{X_\mathrm{surrogate} \to Y}\right),
\label{eq:trE}
\end{equation}

where the second term is the average transfer entropy from $N_{\mathrm{trials}}$ surrogated samples of $X$, to $Y$. 

\subsection{The set $\Gamma$ of most dominant residues}

The total flux between two residues $X$ and $Y$, can be calculated by the equation,

\begin{equation}
D_{X \to Y}=T^E_{X\to Y}-T^E_{Y \to X},
\end{equation}

Residues are selected according to the following rules: $i$ is selected if $D_{X \to Y} > 0$, residue $j$ is selected if $D_{X \to Y} < 0$ and, if $D_{X \to Y}=0$ then no residue is selected. The set of most dominant residues $\Gamma$ is then defined as the set of residues that follow the rules above and also that are above a fixed cutoff $|D_{X \to Y}|\geq D_{\mathrm{cutoff}}$. 

\section{Experiments with GGBP}

\begin{figure}
\includegraphics[width=0.95\linewidth]{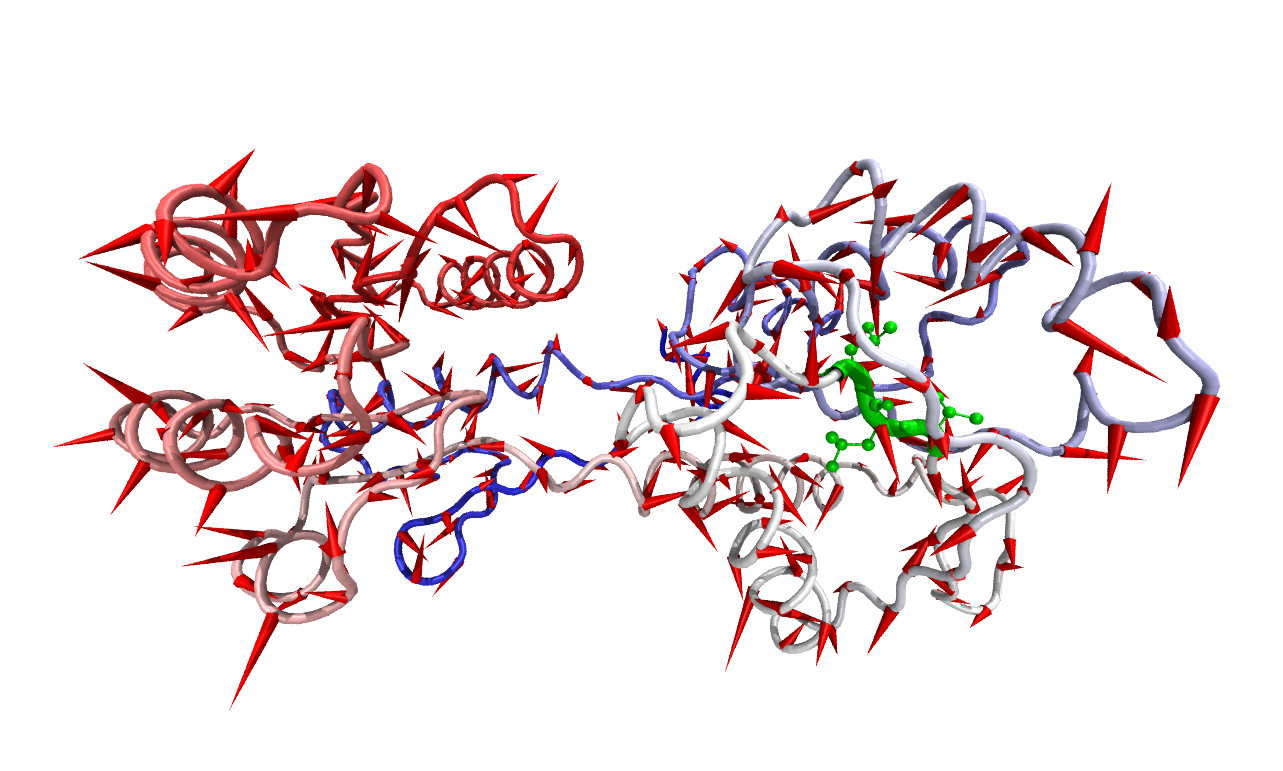}
\caption{A pulling force is applied along the line defined by residues Phe:142 and Leu:144 (highlighted in green). The arrows in red represent the modes determined from a PCA of the pulling trajectory. \label{fig:ggbppull}}
\end{figure}

To verify that our implementation was correct, we performed analysis of coupled chaotic Ulam maps, for Henon maps and for autoregressive processes. In addition, as a more challenging test case, we used the Glucose-galactose binding protein (GGBP) \cite{borrock:2007}. The two domains of GGBP exhibit a 0.5 $rad$ hinge opening motion from one state to the other. The structure of the open state for an unbound glucose-galactose binding protein (GGBP) was crystallized by Borrock et al. (PDBID:2FW0)\cite{borrock:2007} at $1.55$~\AA{}. For the purpose of testing we used both DIMS transitions and we applied a constant pulling force along the line determined by residues Phe:142 and Leu:144 to create a system with a known directional change (highlighted in green in Figure \ref{fig:ggbppull}).

As a comparison point, we performed a PCA analysis over the trajectory generated by pulling along the residues Phe:142 and Leu:144. With the transient nature of the pulling, it can be seen how PCA is unable to detect the pulling direction(Figure \ref{fig:ggbppull}). We now describe the data treatment and some results from our initial testing for the transfer entropy analysis that we propose.

\subsection{Time series treatment}

The time series from MD describing the atomic motions of proteins are real valued. Estimation of the joint probability densities in equation (\ref{eq:trinf}), from real valued data is not only computationally expensive but unnecessary. It has been shown that the amplitude of collective excitations, representing correlated global motions in the protein, samples multicentered distributions \cite{Garcia1992}. Therefore, although single or double precision arithmetic is necessary for the stability and accuracy of the simulations, if done carefully, discretization or coarse-graining of the data does not affect these distributions while greatly reducing noise and increasing computational efficiency. We optimize our implementation by incorporating high performance computing techniques (massively parallel calculations extended over thousands of cores) and by applying dimensionality reduction and data mining techniques that we briefly describe in the following sections.

In other applications of transfer entropies \cite{Kamberaj20091747,Marschinski2002,Gourevitch2007,Lungarella2007,Staniek2008} discretization of the data is performed mainly by using symbolization techniques. In some cases the discretization is so severe that all data is mapped to single bit time series (spikes), as is the case to analyse data from neurophysiological data in epilepsy patients. 

\subsubsection{Piecewise Aggregate Approximation (PAA)}

A time series $\vec{q}(t)=(q_1(t),q_2(t),\ldots,q_{3N}(t)))$ of length $n$ can be represented by a second time series $\vec{Q}(t^\prime)=(Q_1(t^\prime),Q_2(t^\prime),\ldots,Q_{3N}(t^\prime)))$ of length $w<n$, where each element $\vec{Q}(t^\prime)$ is computed according to\cite{Lin2003}:

\begin{equation}
\vec{Q}(t^\prime)=\frac{1}{\Delta t}\int_{t^\prime}^{t^\prime + \Delta t} \vec{q}(t) \mathrm{d} t,
\label{eq:paa}
\end{equation}

where $\Delta t=n/w$. In other words, each vector of the time series $\vec{Q}(t^\prime)$ is simply the average, over a time range $\Delta t$, of the time series $\vec{q}(t)$. When $\Delta t$ is constant, PAA can be seen as an attempt to approximate the original time series with a series of linear functions. Other approaches of PAA include using an adaptive mechanism to adjust $\Delta t$ according to certain rules, i.e. defining a threshold such that $\sigma(t=T) < \langle q(t) - \langle q(t) \rangle_{t=1 \ldots T} \rangle_{t=1 \ldots T} $. For all calculations we set the time range $\Delta t=0.1~$ ns.

\subsection{Transfer entropies from DIMS trajectories}

\begin{figure}
\centering
\includegraphics[width=0.8\linewidth]{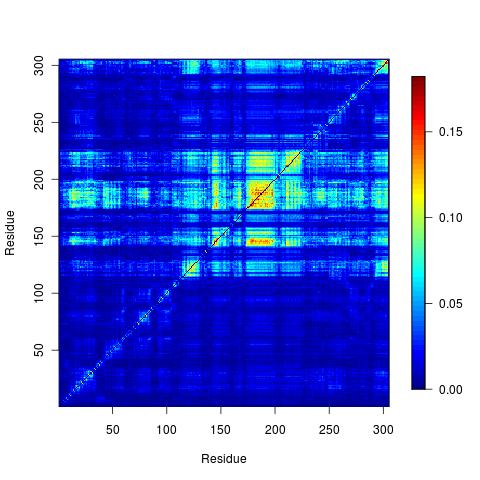}
\caption{Transfer entropies computed for DIMS trajectories for GGBP.\label{fig:dimstrinf}}
\end{figure}

In a previous work we generated a set of transitions for GGBP~\cite{Perilla2010};the simulations were carried out using CHARMM27FF with CMAP ~\cite{charmm:2009} with our implementation of DIMS and using an implicit solvent (ACE2)\cite{schaeferace}. The rotational and translational degrees of freedom were removed by rms fitting the target structure to the evolving system and, the alignment atoms were selected on the N- terminal domain (Residues 111 to 252 and, 293 to 305). By applying our transfer entropy analysis we were able to identify the key residues in the DIMS transition (Figure \ref{fig:dimstrinf}). The results show that the leading residues for the transition are located in the three-segment hinge that connects the N- and C- termini \ref{fig:dimstrinf}. 

\section{Finding the \emph{leading} modes on NtrC}

The structures of the inactive-state and active-state conformations of NtrC have been solved by NMR\cite{Kern1999,Volkman2001,Hastings2003}. At room temperature NtrC samples both conformational states, however after phosphorylation the active states dominate the ensemble set of populations.  Recent studies suggest that the transition pathway between the two conformations can be decomposed in a series of segmented progress variables (order parameters)\cite{Lei2009}. For this study both states were solvated in box of dimensions $20~ \AA{} \times 20~ \AA{} \times 20~ \AA{} $ with TIP3 waters, equilibrated for 15 ns; the total number of atoms, including solvent and ions, is 12168 and 13688 for the active and inactive sates respectively. Production runs were performed for 600 ns using NAMD2.7b2\cite{Phillips2005} at NICS-Kraken. Analysis of the trajectories were executed using our code at NCSA-Abe/Lincoln. 

\subsection{Computing the modes}

A key insight is that the atoms with the strongest leading effective transfer entropy can be used as a subset of degrees of freedom to define collective modes that are new candidate order parameters.  To accomplish this goal, once a cutoff and a time-length for the interrogation of the dynamics has been defined, is straightforward.  The modes are determined by fluctuations of the leading effective transfer components and together describe a set of collective motions.  

\begin{figure}
\centering
\subfigure{\label{fig:involvement1ntr}\includegraphics[width=0.7\columnwidth]{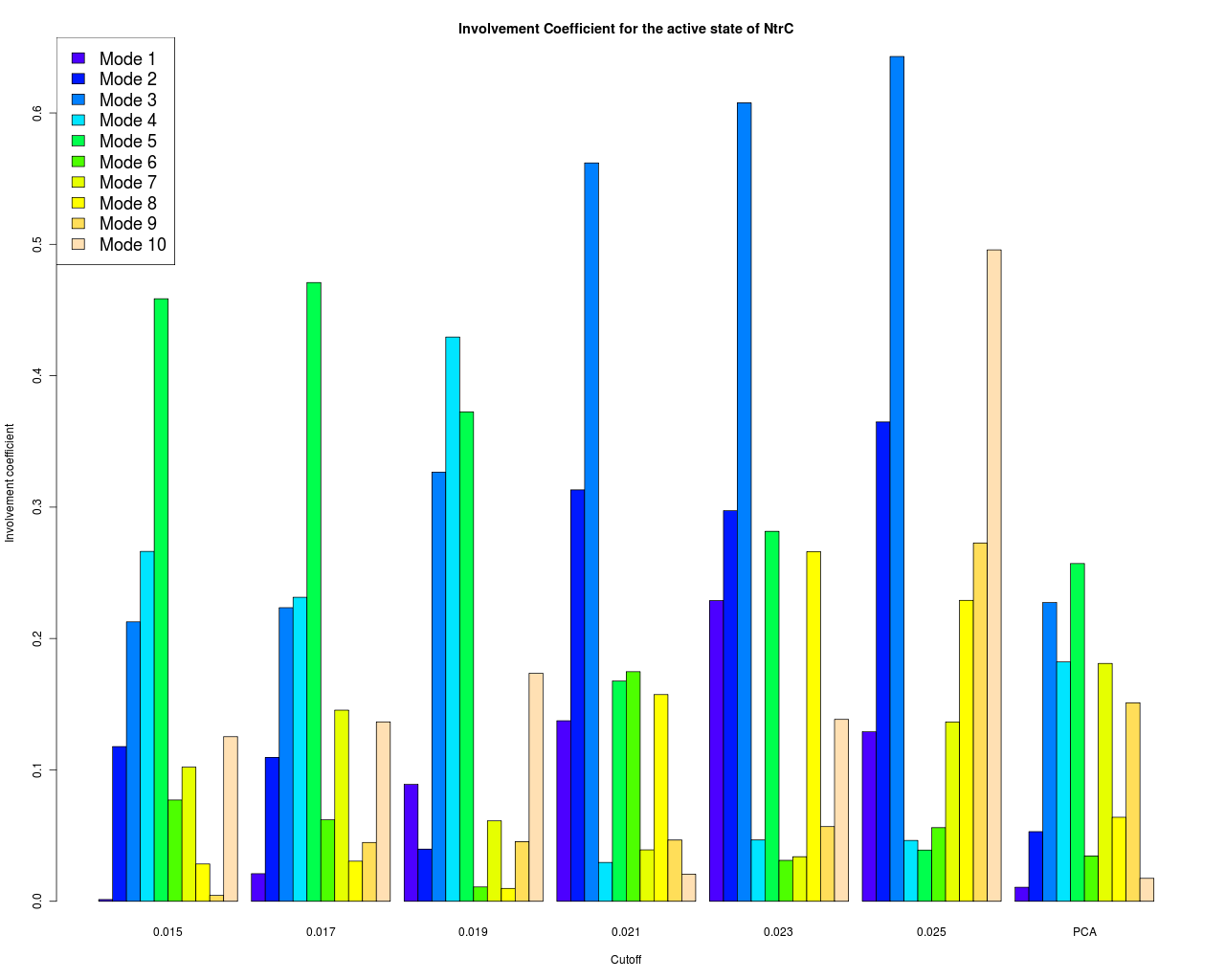}}
\subfigure{\label{fig:involvement1krw}\includegraphics[width=0.7\columnwidth]{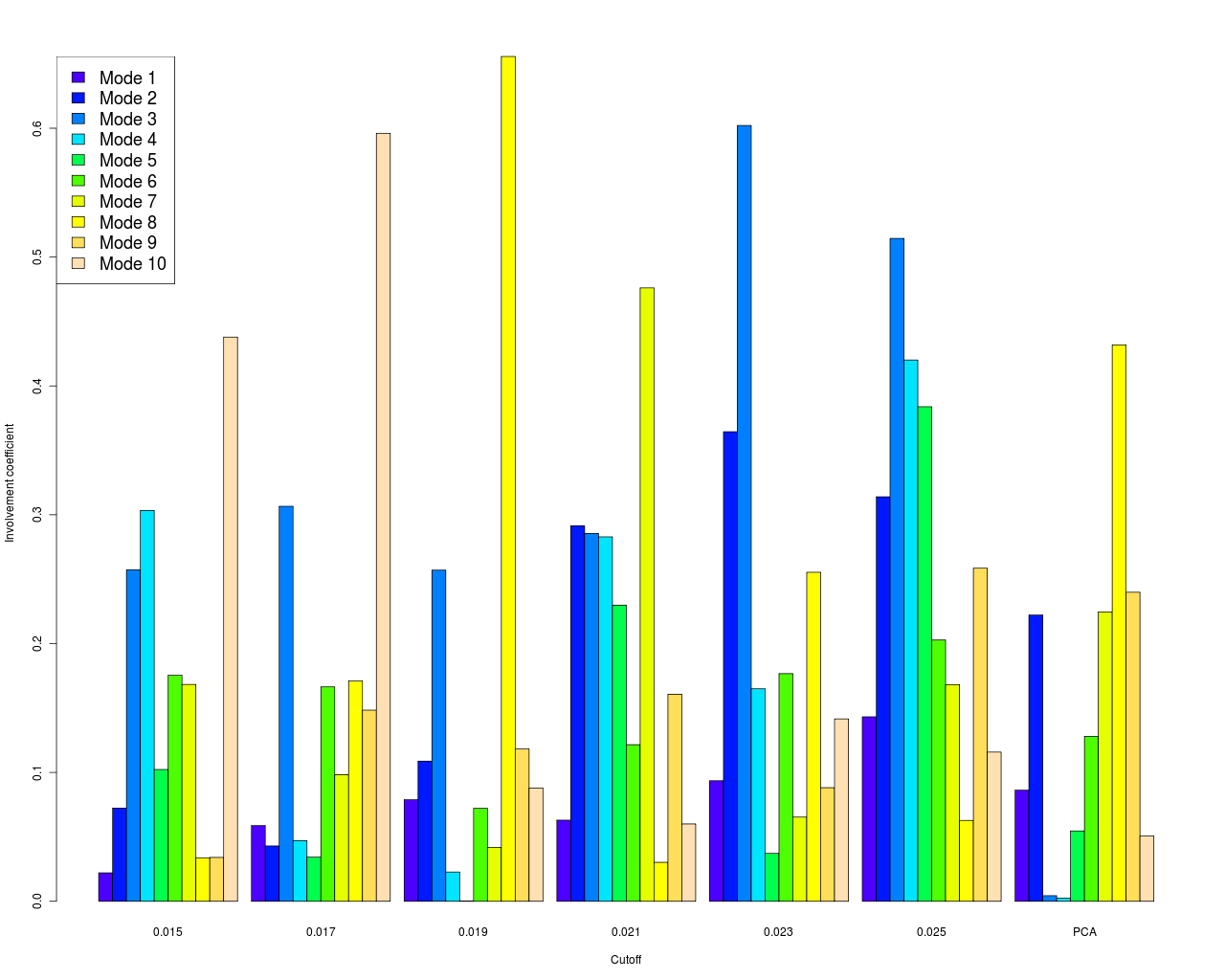}}
\caption{Involvement coefficients for the two states of NtrC for different cutoffs $D_{\mathrm{cutoff}}$.\label{fig:involvement}}
\end{figure}

For the residues in the set $\Gamma$ we compute the correlation matrix as in equation (\ref{eq:cormat}) over the full trajectory and obtain a set of modes $\vec{\eta}_\alpha^{~\prime} $. The involvement coefficients (equation (\ref{eq:involvement})) for different values of the cutoff  $D_{\mathrm{cutoff}}$ are presented in Figure \ref{fig:involvement}. As the cutoff increases fewer residues are selected as dominant, however the involvement coefficients are clearly increasing. This suggests that the most dominant modes $\vec{\eta}_\alpha^{~\prime} $ are pointing towards the end structure. Since the modes are transferring entropy to the entire system biasing along these modes would result in a collective bias for the entire system. 

Since $\eta_\alpha$ is an orthonormal base we can define the cumulative involvement coefficient $\mu_\alpha$ of the first $\alpha$ PCs as:

\begin{equation}
\mu_\alpha = \sum_{i=1}^{\alpha} \nu_i^2,
\end{equation}

and measure how much of the overall difference is accounted by the first $\alpha$ modes. 

\begin{figure}
\centering
\subfigure{\label{fig:cum1ntr}\includegraphics[width=0.7\columnwidth]{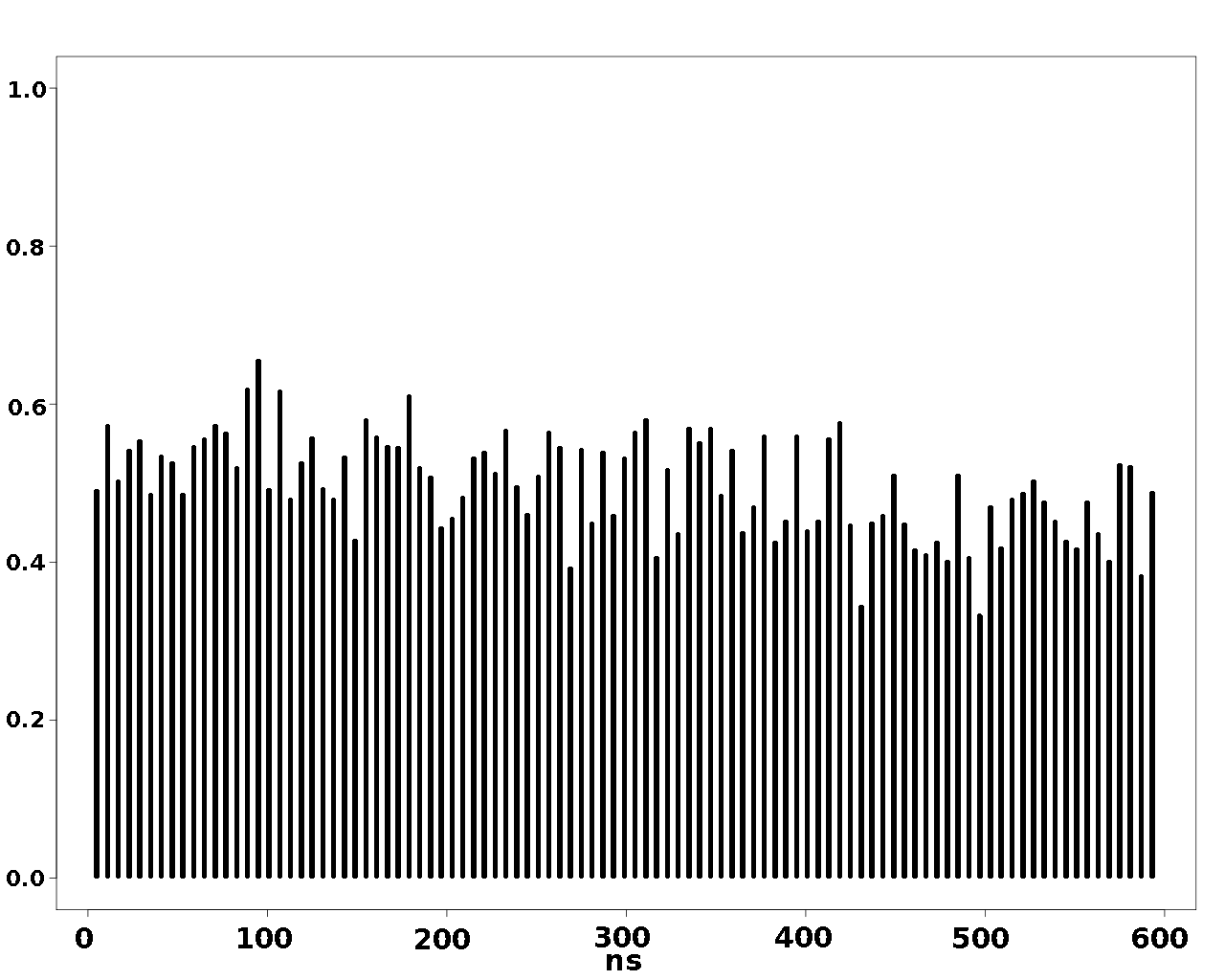}}
\subfigure{\label{fig:cum1krw}\includegraphics[width=0.7\columnwidth]{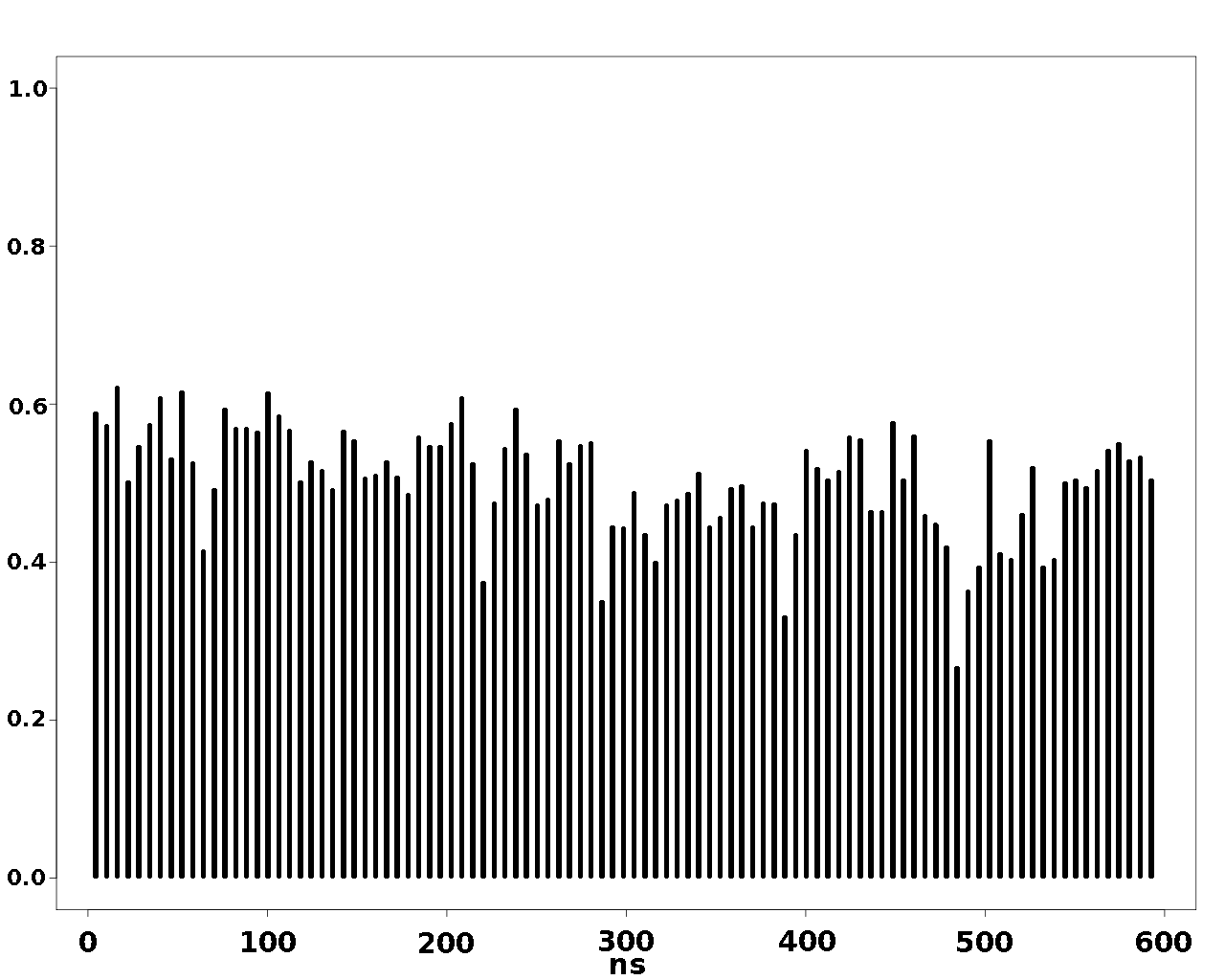}}
\caption{\label{fig:cumminvo}Cumulative involvement coefficient as a function of time(ns) for the first $\alpha=20$ modes.}
\end{figure}

This last figure suggests that relatively short molecular dynamics simulations are converging onto the important degrees of freedom determined by the effective transfer entropy analysis.  It suggests that an algorithm for the use of the effective transfer entropy modes can be readily defined in CHARMM or other computer code.  In that algorithm the lowest frequency modes would be the direction of biasing that is applied through DIMS or another approach (e.g. transition path sampling or targeted MD).  The modes would be defined by a relatively short unbiased simulation and then followed by biasing for a similar amount of time to the mode determination.  For example, this figure would suggest that 5 ns of sampling for the effective modes followed by 5 ns of sampling along the modes could be used to improve the confidence that the most important intermediate states are being reached.  This would then be repeated with unbiased sampling including light restraints on the backbone atoms to define a new set of effective transfer entropy modes.  By continuing this process until the end state is reached, a transition pathway would be defined.  If this process is then repeated for multiple starting points with various sampling windows and different random number seeds, along with a random selection of cutoffs and mode selections, then a good sampling of the intermediate space should be obtained.  We expect to implement this type of algorithm in the future and the results will be presented at that time.





\section{Conclusions}
 
A molecular understanding of how protein function is related to protein structure will require an ability to understand large conformational changes between multiple states.  Unfortunately these states are often separated by high free energy barriers and within a complex energy landscape. This makes it very difficult to reliably connect, for example, by all-atom molecular dynamics calculations, the states, their energies and the pathways between them.  A major issue needed to improve sampling on the intermediate states is an order parameter -- a reduced descriptor for the major subset of degrees of freedom -- that can be used to aid sampling for the large conformational change.  In this paper we present a novel way to combine information from molecular dynamics using non-linear time series and dimensionality reduction, in order to quantitatively determine an order parameter connecting two large-scale conformationally distinct protein states. The results presented show that the leading modes can be computed from short simulations. This new method suggests an implementation for molecular dynamics calculations that may dramatically enhance sampling of intermediate states.


%

\end{document}